\documentclass[conference]{IEEEtran}
\IEEEoverridecommandlockouts
\usepackage{cite}
\usepackage{amsmath,amssymb,amsfonts}
\usepackage{algorithmic}
\usepackage{graphicx}
\usepackage{textcomp}
\usepackage{xcolor}
\def\BibTeX{{\rm B\kern-.05em{\sc i\kern-.025em b}\kern-.08em
    T\kern-.1667em\lower.7ex\hbox{E}\kern-.125emX}}
\usepackage[colorlinks=true, allcolors=blue]{hyperref}
\usepackage{multirow}
\newcommand{\comment}[1]{}
\begin{document}

\title{On the application of matrix congruence to QUBO formulations for systems of linear equations\\
\thanks{This work was partially supported by National Institute for Mathematical Sciences (NIMS) grant funded by the Korean Government (MSIT) B21810000.}
}

\author{\IEEEauthorblockN{Sun Woo Park}
\IEEEauthorblockA{\textit{Department of Mathematics} \\
\textit{University of Wisconsin-Madison}\\
Madison, WI. USA. \\
\textit{\; \; National Institute for Mathematical Sciences \; \;}\\
Daejon, Republic of Korea.\\
spark483@wisc.edu / spark483@nims.re.kr}
\and
\IEEEauthorblockN{Hyunju Lee}
\IEEEauthorblockA{
\textit{\; \; Innovation on Quantum and Computed Tomography Inc.}\\
Seoul, Republic of Korea.\\
hjlee@iqct.io}
\and
\IEEEauthorblockN{Byung Chun Kim}
\IEEEauthorblockA{
\textit{Innovation on Quantum and Computed Tomography Inc.}\\
Seoul, Republic of Korea.\\
bckim@iqct.io}
\and
\IEEEauthorblockN{Youngho Woo}
\IEEEauthorblockA{
\textit{\; \; National Institute for Mathematical Sciences \; \; \; \;}\\
Daejon, Republic of Korea.\\
youngw@nims.re.kr}
\and
\IEEEauthorblockN{Kyungtaek Jun**}
\IEEEauthorblockA{
\textit{Innovation on Quantum and Computed Tomography Inc.}\\
Seoul, Republic of Korea.\\
ktfriends@gmail.com}
}

\maketitle

\begin{abstract}
Recent studies on quantum computing algorithms focus on excavating features of quantum computers which have potential for contributing to computational model enhancements. Among various approaches, quantum annealing methods effectively parallelize quadratic unconstrained binary optimization (QUBO) formulations of systems of linear equations. In this paper, we simplify these formulations by exploiting congruence of real symmetric matrices to diagonal matrices. We further exhibit computational merits of the proposed QUBO models, which can outperform classical algorithms such as QR and SVD decomposition.
\end{abstract}

** Corresponding Author: Kyungtaek Jun.\\
Sun Woo Park and Hyunju Lee made equal contributions as co-first authors.

\section{INTRODUCTION}
\label{sec:intro}  % \label{} allows reference to this section
Quantum computing methods have been gaining attention as a potential candidate for effectively enhancing pre-existent classical algorithms. As one of such computing techniques, quantum annealing method focuses on returning solutions which minimize energy level functions associated to optimization problems \cite{RI14, MV16}. The problem of finding a solution to a system of linear equations can be reformulated as an optimization problem. Borle and Lomonaco proposed a quadratic unconstrained binary optimization (QUBO) formulation of solving systems of linear equations \cite{BL19}. Using the binary expansions of real numbers, quantum annealing processors effectively parallelize the process of evaluating the minimum energy level of the QUBO model. As a result, quantum annealing based QUBO model is computationally cost efficient compared to classical algorithms such as QR and SVD. However, physical conditions arising from using quantum processors, such as the required number of qubits for the model, may negatively affect the accuracy of such models \cite{BL19, JC21}. It is thus a natural question to ask what additional measures can further enhance the accuracy of quantum annealing based QUBO models.

%%% Hamiltonian matrix가 아니예용 ㅠㅠ energy Hamiltonian으로 부터 만들어지는 matrix라고 합니다. 처음에 만든애가 이름을 안넣었어요 ㅠㅠ 그리고 헤밀토니안 메이트릭스는 수학적으로 다른게 또 있어요! 수정이 필요합니다 ^^
%%%\textcolor{red}{Fix Hamiltonian matrix}\\
In this paper, we propose a new simplified QUBO formulation of systems of linear equations which utilizes a classical result that real symmetric matrices are congruent to real diagonal matrices \cite{Ay62}. We show that, under mild conditions, such a transformation substantially reduces the number of non-trivial relations among qubits used in formulating the QUBO model. Furthermore, we demonstrate with an example that utilizing the classical result greatly enhances the accuracy of quantum annealing based QUBO models.

\section{Method}
\label{sec: method}

\subsection{Background}
\label{subsec: background}

Let $A := (a_{i,j})_{i,j=1}^n \in \mathbb{R}^{n \times n}$ be an $n \times n$ matrix, and $b := (b_i)_{i=1}^n \in \mathbb{R}^n$ an arbitrary column vector of dimension $n$. Denote by $x := (x_i)_{i=1}^n \in \mathbb{R}^n$ the column vector of $n$ variables which satisfies
\begin{equation} \label{equation:linear_system}
    Ax = b.
\end{equation}
The solution to the systems of linear equations specified in (\ref{equation:linear_system}) is the $l^2$-norm minimizing solution of 
\begin{equation} \label{equation:minimize_square}
    \| Ax - b \|^2 = x^T A^T A x - 2b^T A x + b^T b.
\end{equation}
Because $A^T A$ is a positive semidefinite symmetric matrix over $\mathbb{R}$, it is congruent to a diagonal matrix $D \in \mathbb{R}^{n \times n}$ \cite{Ay62}. In particular, there exists a non-singular matrix $R := (r_{i,j})_{i,j=1}^n \in \mathbb{R}^{n \times n}$ such that $R^T (A^T A) R$ is a diagonal matrix. Denote by $D := (d_{i,j})_{i,j=1}^n \in \mathbb{R}^{n \times n}$ the diagonal matrix $R^T (A^T A) R$. 

We can hence write the $l^2$-norm of $Ax - b$ as:
\begin{equation} \label{equation:sylvester_1}
    \| Ax - b \|^2 = (R^{-1} x)^T R^T A^T A R  (R^{-1} x) - 2b^T A R (R^{-1} x) + b^T b
\end{equation}
Because $R$ is invertible, there exists a column vector $y \in \mathbb{R}^n$ such that $y = R^{-1} x.$ Substituting $y$ to (\ref{equation:sylvester_1}) gives:
\begin{equation} \label{equation:sylvester_2}
    \|Ax - b\|^2 = y^T D y - 2 (b^T A R) y + b^T b.
\end{equation}
Up to linear change of variables, the solution to (\ref{equation:linear_system}) is the $l^2$-norm minimizing solution of (\ref{equation:sylvester_2}).
\begin{align} \label{equation:sylvester_final}
    \|Ax - b\|^2 &= \sum_{i=1}^n d_{i,i} y_i^2 - 2 \sum_{i=1}^n \sum_{j=1}^n \sum_{k=1}^n  b_k a_{k,j} r_{j,i} y_i + \sum_{i=1}^n b_i^2
\end{align}

In the spirit of quantum annealing approach \cite{MV16, BL19, JC21, J21} to formulating systems of linear equations, the $l^2$-norm minimizing solution of (\ref{equation:sylvester_2}) can be approximately represented by a combination of qubits $q_{i,j} \in \{0,1\}$. Throughout this paper, we assume that the variables are approximated by the radix 2 representation:
%%%\textcolor{red}{Do you have any region to use $m_i$-1? I think $m$ is the best because you used  the range of m: $-m <= l1, l2 <= m$}

\begin{equation} \label{equation:QUBO_model_1}
    y_i \approx \sum_{l=-m}^{m} 2^{l} q_{i,l}^+ - \sum_{l=-m}^{m} 2^{l} q_{i,l}^-.
\end{equation}
We denote by $m$ the upper bound on the number of digits or fractional digits used to represent $y_i$. That is, we allow $l$, a digit or a fractional digit of $y_i$, to take values between $-m$ and $m$.  Note that the qubits $q_{i,l}^\pm$ can be chosen such that for any digits $-m \leq l_1, l_2 \leq m$, $q_{i,l_1}^+ q_{i,l_2}^- = 0$ \cite{J21}.

\subsection{Example}

We demonstrate with an example to show how using matrix congruence relation significantly simplifies the QUBO model associated to the system of linear equations. Consider the following system of linear equations $Ax = b$ \cite{JC21}:
\begin{equation} \label{equation:sample_example}
    \begin{pmatrix}
    3 & 1 \\
    -1 & 2
    \end{pmatrix}
    \begin{pmatrix}
    x_1 \\
    x_2
    \end{pmatrix}
    =
    \begin{pmatrix}
    -1 \\
    5
    \end{pmatrix}.
\end{equation}
Suppose that the matrices $D$ and $R$ associated to $A^T A$ are given.
\begin{equation}
    D = \begin{pmatrix}
    \frac{8}{5} & 0 \\ 0 & \frac{98}{125}
    \end{pmatrix}, \; \; 
    R = \begin{pmatrix}
    \frac{2}{5} & -\frac{1}{25} \\
    0 & \frac{2}{5}
    \end{pmatrix}.
\end{equation}
We use the radix 2 representation of the column vector $y = R^{-1}x$:
\begin{equation} \label{equation:example_QAOA}
    \begin{pmatrix}
    y_1 \\ y_2
    \end{pmatrix}
    =
    \begin{pmatrix}
    q_{11} + 2q_{12} + 4q_{13} - q_{14} - 2q_{15} - 4q_{16} \\ q_{21} + 2q_{22} + 4q_{23} - q_{24} - 2q_{25} - 4q_{26}
    \end{pmatrix}
\end{equation}
Define the cost function for solving the system of linear equations as
\begin{align} \label{equation:sylvester_final_1}
    f(y) := \sum_{i=1}^2 d_{i,i} y_i^2 - 2 \sum_{i=1}^2 \sum_{j=1}^2 \sum_{k=1}^2 b_k a_{k,j} r_{j,i} y_i,
\end{align}
where our objective is to find a column vector $y \in \mathbb{R}^n$ such that $f(y) = -b^T b = -26$. Observe that (\ref{equation:sylvester_final_1}) is the expansion of the first two terms of (\ref{equation:sylvester_final}). We recall the following conditions for $i \in \{1,2\}$ and $j \in \{1,2,\cdots,6\}$:
\begin{align} \label{equation:equivalence_relation}
\begin{cases}
    & q_{i,1} q_{i,4} = q_{i,2} q_{i,4} = q_{i,3} q_{i,4} = 0 \\
    & q_{i,1} q_{i,5} = q_{i,2} q_{i,5} = q_{i,3} q_{i,5} = 0 \\
    & q_{i,1} q_{i,6} = q_{i,2} q_{i,6} = q_{i,3} q_{i,6} = 0 \\
    & q_{i,j}^2 = q_{i,j}
\end{cases}
\end{align}
The first three conditions use the fact that for any digits $0 \leq l_1, l_2 \leq 2$, $q_{i,l_1}^+ q_{i,l_2}^- = 0$. The last condition holds because each qubit takes values of either $0$ or $1$.

Substituting (\ref{equation:example_QAOA}) to (\ref{equation:sylvester_final_1}) under the aforementioned conditions from (\ref{equation:equivalence_relation}) yields:
%%%\textcolor{red}{Can we delete all squares?}
\begin{align}
\begin{split}
    f(y) &= 8q_{11} + \frac{32}{5}q_{11}q_{12} + \frac{64}{5}q_{11}q_{13} + \frac{96}{5}q_{12} + \frac{128}{5}q_{12}q_{13} \\
    &+ \frac{256}{5}q_{13} - \frac{24}{5}q_{14} + \frac{32}{5}q_{14}q_{15} + \frac{64}{5}q_{14}q_{16} - \frac{32}{5}q_{15} \\
    &+  \frac{128}{5}q_{15}q_{16} - \frac{882}{125}q_{21} + \frac{392}{125}q_{21}q_{22} - \frac{1568}{125}q_{22} \\
    &+ \frac{784}{125}q_{21}q_{23} + \frac{1568}{125}q_{22}q_{23} - \frac{2352}{125}q_{23} + \frac{1078}{125}q_{24} \\
    &+ \frac{392}{125}q_{24}q_{25} + \frac{2352}{125}q_{25} + \frac{784}{125}q_{24}q_{26} + \frac{1568}{125}q_{25}q_{26} \\
    &+ \frac{5488}{125}q_{26}.
\end{split}
\end{align}

\begin{figure*}[!t]
% ensure that we have normalsize text
\normalsize
% Store the current equation number.
\setcounter{equation}{12}

\begin{equation}\label{equation:Qhat_y}
{\small
    \hat{Q} :=
    \begin{pmatrix}
    8 & 6.4 & 12.8  & 0  & 0 & 0 & 0 & 0 & 0 & 0 & 0 & 0 \\
    0  & 19.2  & 25.6  & 0  & 0 & 0 & 0 & 0 & 0 & 0 & 0 & 0 \\
    0  & 0  & 51.2 & 0 & 0 & 0 & 0 & 0 & 0 & 0 & 0 & 0 \\
    0  & 0  & 0  & -4.8 & 6.4 & 12.8 & 0 & 0 & 0 & 0 & 0 & 0 \\
    0  & 0  & 0  & 0  & -6.4 & 25.6 & 0 & 0 & 0 & 0 & 0 & 0 \\
    0  & 0  & 0  & 0  & 0  &  0 & 0 & 0 & 0 & 0 & 0 & 0 \\
    0  & 0  & 0  & 0  & 0  &  0 & -7.056 & 3.136 & 6.272 & 0 & 0 & 0 \\
    0  & 0  & 0  & 0  & 0  &  0 & 0 & -12.544 & 12.544 & 0 & 0 & 0 \\
    0  & 0  & 0  & 0  & 0  &  0 & 0 & 0 & -18.816 & 0 & 0 & 0 \\
    0  & 0  & 0  & 0  & 0  &  0 & 0 & 0 & 0 & 8.624 & 3.136 & 6.272 \\
    0  & 0  & 0  & 0  & 0  &  0 & 0 & 0 & 0 & 0 & 18.816 & 12.544 \\
    0  & 0  & 0  & 0  & 0  &  0 & 0 & 0 & 0 & 0 & 0 & 43.904 \\
    \end{pmatrix},
}
\end{equation}

\begin{equation}\label{equation:Qhat_prime_y}
{\small
    \hat{Q}' :=
    \begin{pmatrix}
    26 & 40 & 80  & -20  & -40 & -80 & 2 & 4 & 8 & -2 & -4 & -8 \\
    0  & 72  & 160  & -40  & -80 & -160 & 4 & 8 & 16 & -4 & -8 & -16 \\
    0  & 0  & 224 & -80 & -160 & -320 & 8 & 16 & 32 & -8 & -16 & -32 \\
    0  & 0  & 0  & -6 & 40 & 80 & -2 & -4 & -8 & 2 & 4 & 8 \\
    0  & 0  & 0  & 0  & 8 & 160 & -4 & -8 & -16 & 4 & 8 & 16 \\
    0  & 0  & 0  & 0  & 0  &  96 & -8 & -16 & -32 & 8 & 16 & 32 \\
    0  & 0  & 0  & 0  & 0  &  0 & -13 & 20 & 40 & -10 & -20 & -40 \\
    0  & 0  & 0  & 0  & 0  &  0 & 0 & -16 & 80 & -20 & -40 & -80 \\
    0  & 0  & 0  & 0  & 0  &  0 & 0 & 0 & 8 & -40 & -80 & -160 \\
    0  & 0  & 0  & 0  & 0  &  0 & 0 & 0 & 0 & 23 & 20 & 40 \\
    0  & 0  & 0  & 0  & 0  &  0 & 0 & 0 & 0 & 0 & 56 & 80 \\
    0  & 0  & 0  & 0  & 0  &  0 & 0 & 0 & 0 & 0 & 0 & 152 \\
    \end{pmatrix},
}
\end{equation}
% IEEE uses as a separator
\hrulefill
% The spacer can be tweaked to stop underfull vboxes.
\vspace*{4pt}
\end{figure*}
Let $\hat{Q}$ be the matrix defined as in (\ref{equation:Qhat_y}). Denote by $q_y := [q_{11}, q_{12}, \cdots q_{26}]^T$ the column vector of qubits used in the radix 2 representations of $y_1$ and $y_2$. Then the energy function $f(y)$ satisfies 
\begin{equation} \label{equation:new_qubo_model}
    f(y) = q_y^T \hat{Q} q_y = y^T D y - 2 b^T A R y,
\end{equation}
up to the equivalence relation $q_{i,j}^2 = q_{i,j}$. We note that the matrix $\hat{Q}$ characterizes the inherent relations among the qubits used in representing the variables $y_i$. Solving the system of linear equations $Ax = b$ is thus equivalent to finding the column vector $q_y$ such that $q_y^T \hat{Q} q_y = b^T b = -26$.
%%%\textcolor{red}{In this paper, we are using y as a variable. So, how about using another y that satisfies bt*b as a solution.}\\

Suppose, on the other hand, we use the radix 2 representation of the column vector $x$ from (\ref{equation:minimize_square}) \cite{JC21}. As before, denote by $q_x := [q_{11},q_{12},\cdots,q_{26}]^T$ the column vector of qubits used in the radix 2 representations of $x_1$ and $x_2$. Let $\hat{Q}'$ be the matrix defined as in (\ref{equation:Qhat_prime_y}).
%\begin{equation}\label{equation:Qhat_x}
%    \hat{Q}' :=
%    \begin{pmatrix}
%    26 & 40 & -20  & -40  & 2 & 4 & -2 & -4 \\
%    0  & 72  & -40  & -80  & 4 & 8 & -4 & -8 \\
%    0  & 0  & -6 & 40 & -2 & -4 & 2 & 4 \\
%    0  & 0  & 0  & 8 & -4 & -8 & 4 & 8 \\
%    0  & 0  & 0  & 0  & -13 & 20 & -10 & -20 \\
%    0  & 0  & 0  & 0  & 0   & -16 & -20 & -40 \\
%    0  & 0  & 0  & 0  & 0   & 0    & 23 & 20 \\
%    0  & 0  & 0  & 0  & 0 & 0 & 0 & 56
%    \end{pmatrix},
%\end{equation}
Then, up to the equivalence relation $q_{i,j}^2 = q_{i,j}$, we have 
\begin{equation} \label{equation:vanilla_qubo_model}
    q_x^T \hat{Q}' q_x = x^T A^T A x - 2 b^T A x
\end{equation}
The matrix $\hat{Q}'$ characterizes the inherent relations among the qubits used in representing the variables $x_i$. One can solve $Ax = b$ by finding $q_x$ that satisfies $q_x^T \hat{Q}' q_x = -26$.

\section{Implementation}
We implement the aforementioned example (\ref{equation:sample_example}) on D-Wave 2000Q quantum annealer. Both QUBO models obtained from (\ref{equation:new_qubo_model}) and (\ref{equation:vanilla_qubo_model}) are performed for 3 trials with 10,000 anneals. 

\subsection{Vanilla QUBO model}
Without using matrix congruence, the solution to the 2-dimensional linear systems of equation is given by $x_1 = -1$ and $x_2 = 2$. The vanilla QUBO model (\ref{equation:vanilla_qubo_model}) aims to search for possible combinations of qubits $q_x$. We list all possible combinations of qubits for $x_i = q_{i1} + 2q_{i2} + 4q_{i3} - q_{i4} - 2q_{i5} - 4q_{i6}$ in (\ref{equation:qubits_all_combinations}). There are 7 possible combinations for $x_1 = -1$, and 6 possible combinations for $x_2 = 2$.
\begin{align} \label{equation:qubits_all_combinations}
    \begin{split}
        (q_{11}, q_{12}, q_{13}, q_{14}, q_{15}, q_{16}) \in \{ &(0,0,0,1,0,0),(0,1,0,1,1,0), \\ 
        &(0,0,1,1,0,1),(0,1,1,1,1,1), \\
        &(1,0,0,0,1,0),(1,0,1,0,1,1), \\
        &(1,1,0,0,0,1) \} \\
        (q_{21},q_{22},q_{23},q_{24},q_{25},q_{26}) \in \{ &(0,0,1,0,1,0),(0,1,0,0,0,0),\\
        &(0,1,1,0,0,1),(1,0,1,1,1,0),\\
        &(1,1,0,1,0,0),(1,1,1,1,0,1) \}
    \end{split}
\end{align}

%%%\begin{table}[hbt!]
%\begin{center}
%{\small
%\begin{tabular}{c|c|c|c|c|c||c|c|c|c|c|c}
%\hline \hline
%\multicolumn{6}{c||}{$x_1 = -1$} & \multicolumn{6}{c}{$x_2 = 2$} \\ \hline
%$q_{11}$ & $q_{12}$ & $q_{13}$ & $q_{14}$ & $q_{15}$ & $q_{16}$ & $q_{21}$ & $q_{22}$ & $q_{23}$ & $q_{24}$ & $q_{25}$ & $q_{26}$ \\ \hline \hline
%0 & 0 & 0 & 1 & 0 & 0 & 0 & 0 & 1 & 0 & 1 & 0 \\ \hline
%0 & 1 & 0 & 1 & 1 & 0 & 0 & 1 & 0 & 0 & 0 & 0 \\ \hline
%0 & 0 & 1 & 1 & 0 & 1 & 0 & 1 & 1 & 0 & 0 & 1 \\ \hline
%0 & 1 & 1 & 1 & 1 & 1 & 1 & 0 & 1 & 1 & 1 & 0 \\ \hline
%1 & 0 & 0 & 0 & 1 & 0 & 1 & 1 & 0 & 1 & 0 & 0 \\ \hline
%1 & 0 & 1 & 0 & 1 & 1 & 1 & 1 & 1 & 1 & 0 & 1 \\ \hline
%1 & 1 & 0 & 0 & 0 & 1 & \multicolumn{6}{c}{} \\ \hline \hline
%\end{tabular}
%}
%\end{center}
%\caption{A list of all possible qubit combinations of the QUBO model from (\ref{equation:Qhat_prime_y})}
%\label{tab:original_QUBO_qubit}
%\end{table}

Table \ref{tab:original_QUBO} lists the results of performing the vanilla QUBO model for 3 trials on the D-Wave quantum annealer with 10,000 anneals. Here, we abbreviated the 6 possible combinations of qubits for $x_2 = 2$. Each row lists the number of occurrences with the lowest energy with the given combination of qubits $q_{11}, q_{12}, \cdots, q_{16}$ for $x_1 = -1$. Out of 3 trials, the D-Wave quantum annealer finds 887, 1181, and 1065 occurrences with the lowest energy out of 10,000 anneals.

\begin{table*}[htbp]
\begin{center}
{\small
\begin{tabular}{c|c|c|c|c|c||c|c|c|c|c|c||c||c|c|c}
\hline
\hline
\multirow{2}{*}{$q_{11}$} & \multirow{2}{*}{$q_{12}$} & \multirow{2}{*}{$q_{13}$} & \multirow{2}{*}{$q_{14}$} & \multirow{2}{*}{$q_{15}$} & \multirow{2}{*}{$q_{16}$} & \multirow{2}{*}{$q_{21}$} & \multirow{2}{*}{$q_{22}$} & \multirow{2}{*}{$q_{23}$} & \multirow{2}{*}{$q_{24}$} & \multirow{2}{*}{$q_{25}$} & \multirow{2}{*}{$q_{26}$} & \multirow{2}{*}{Energy} & \multicolumn{3}{c}{\# Occurrences} \\ \cline{14-16} & & & & & & & & & & & & & Run 1 & Run 2 & Run 3 \\ \hline \hline
0 & 0 & 0 & 1 & 0 & 0 & \multicolumn{6}{c||}{All 6 combinations} & -26.0 & 203 & 66 & 50 \\ \hline
0 & 1 & 0 & 1 & 1 & 0 & \multicolumn{6}{c||}{All 6 combinations} & -26.0 & 77 & 49 & 531 \\ \hline
0 & 0 & 1 & 1 & 0 & 1 & \multicolumn{6}{c||}{All 6 combinations} & -26.0 & 131 & 147 & 251 \\ \hline 
0 & 1 & 1 & 1 & 1 & 1 & \multicolumn{6}{c||}{All 6 combinations} & -26.0 & 71 & 116 & 51 \\ \hline
1 & 0 & 0 & 0 & 1 & 0 & \multicolumn{6}{c||}{All 6 combinations} & -26.0 & 75 & 43 & 74 \\ \hline 
1 & 0 & 1 & 0 & 1 & 1 & \multicolumn{6}{c||}{All 6 combinations} & -26.0 & 71 & 83 & 62 \\ \hline
1 & 1 & 0 & 0 & 0 & 1 & \multicolumn{6}{c||}{All 6 combinations} & -26.0 & 259 & 677 & 46 \\ \hline
\hline 
\multicolumn{11}{c}{} & & Total & 887 & 1181 & 1065 \\ \hline
\hline
\end{tabular}
}
\end{center}
\caption{Number of occurrences with the lowest energy levels using the QUBO model from (\ref{equation:Qhat_prime_y})}
\label{tab:original_QUBO}
\end{table*}

\subsection{New QUBO model using matrix congruences}
Using matrix congruence, the solution to (\ref{equation:sample_example}) is given by $y_1 = -2$ and $y_2 = 5$. Because we further simplified the QUBO model using the equivalence relation, as shown in (\ref{equation:equivalence_relation}), the new QUBO model (\ref{equation:new_qubo_model}) searches for the following unique combination of qubits $y_i = q_{i1} + 2q_{i2} + 4q_{i3} - q_{i4} - 2q_{i5} - 4q_{i6}$:
\begin{align}
    \begin{split}
        (q_{11}, q_{12}, q_{13}, q_{14}, q_{15}, q_{16}) &= (0, 0, 0, 0, 1, 0) \\
        (q_{21}, q_{22}, q_{23}, q_{24}, q_{25}, q_{26}) &= (1, 0, 1, 0, 0, 0) 
    \end{split}
\end{align}
To check whether specifying the zero terms of the matrix $\hat{Q}$ from (\ref{equation:Qhat_y}) to the D-Wave system affects the performance of the linear system solver algorithm, we implement the new QUBO model under two different conditions. The first method specifies the zero terms of the upper triangular portion of $\hat{Q}$ in the D-Wave systems code, whereas the second method omits these zero terms from the code. Table \ref{tab:new_QUBO_with_0} displays the number of occurrences of the unique combination of qubits out of 10,000 anneals. The former method finds 1526, 2495, and 2063 occurrences out of three runs, whereas the latter method finds 2103, 4441, and 1727 occurrences.

\begin{table*}[]
\centering
{\small
\begin{tabular}{c|c|c|c|c|c||c|c|c|c|c|c||c|c||c|c|c}
\hline
\hline
\multirow{2}{*}{$q_{11}$} & \multirow{2}{*}{$q_{12}$} & \multirow{2}{*}{$q_{13}$} & \multirow{2}{*}{$q_{14}$} & \multirow{2}{*}{$q_{15}$} & \multirow{2}{*}{$q_{16}$} & \multirow{2}{*}{$q_{21}$} & \multirow{2}{*}{$q_{22}$} & \multirow{2}{*}{$q_{23}$} & \multirow{2}{*}{$q_{24}$} & \multirow{2}{*}{$q_{25}$} & \multirow{2}{*}{$q_{26}$} & \multirow{2}{*}{Energy} & \multirow{2}{*}{Zero Terms} & \multicolumn{3}{c}{\# Occurrences} \\ \cline{15-17} & & & & & & & & & & & & & & Run 1 & Run 2 & Run 3 \\ \hline \hline
0 & 0 & 0 & 0 & 1 & 0 & 1 & 0 & 1 & 0 & 0 & 0 & -26.0 & Yes & 1526 & 2495 & 2063 \\ \hline
0 & 0 & 0 & 0 & 1 & 0 & 1 & 0 & 1 & 0 & 0 & 0 & -26.0 & No & 2103 & 4441 & 1727 \\ \hline
\hline 
\end{tabular}
}
\centering
\caption{Number of occurrences with the lowest energy levels using the QUBO model from (\ref{equation:Qhat_y}). The first row shows the case where the zero terms are specified in the D-Wave code. The second row displays the case where the zero terms are omitted from the code.}
\label{tab:new_QUBO_with_0}
\end{table*}

\section{Discussion}
The new QUBO model (\ref{equation:new_qubo_model}), regardless of whether the zero terms of $\hat{Q}$ are specified in the D-Wave code, clearly outperforms the vanilla QUBO model (\ref{equation:vanilla_qubo_model}). A summary of the number of occurrences with the lowest energy levels using three QUBO model implementations is demonstrated in Table \ref{tab:Results Comparison}. 

\begin{table*}[hbt!]
\centering
{\small
\begin{tabular}{c||c|c|c}
\hline
\hline
\multirow{2}{*}{\# Trial} & Vanilla QUBO model & New QUBO model & New QUBO model \\ 
& (Table \ref{tab:original_QUBO}) & (Table \ref{tab:new_QUBO_with_0}, row 1) & (Table \ref{tab:new_QUBO_with_0}, row 2) \\ \hline
\hline
Run 1 & 887 & 1526 & 2103 \\ \hline
Run 2 & 1181 & 2495 & 4441 \\ \hline
Run 3 & 1065 & 2063 & 1727 \\ \hline
\hline
Average \# Occurrences & 1044 & 2028 & 2758 \\ \hline
Average Probability & 10.44\% & 20.28 \% & 27.58 \% \\
\hline
\hline
\end{tabular}
}
\centering
\caption{A summary of the number of occurrences with the lowest energy levels for each QUBO model}
\label{tab:Results Comparison}
\end{table*}

On average, the probability that the new QUBO model solves (\ref{equation:sample_example}) ranges between $20.28 \%$ and $27.58\%$. Compare this to $10.44\%$, the average probability obtained from the vanilla QUBO model. This is roughly half of the probability obtained from the proposed model. We also observe that omitting the zero entries from the implementation further enhances the performance of the new QUBO model. 

The outperformance of the new QUBO model can be traced from the block diagonalization of the matrix $Q$ characterizing the relations among qubits used in the model, as constructed in (\ref{equation:Qhat_y}),(\ref{equation:new_qubo_model}),(\ref{equation:Qhat_prime_y}), and (\ref{equation:vanilla_qubo_model}). The example (\ref{equation:sample_example}) is a system of $2$ linear equations with $2$ variables. As shown in (\ref{equation:example_QAOA}), the two unknown variables are approximated by the radix 2 representation with $3$ digits, using a total of $12$ qubits. The two matrices $\hat{Q}$ and $\hat{Q}'$, both characterizing the respective QUBO models, are $12 \times 12$ matrices containing the relations among 12 qubits. Consider the matrix $\hat{Q}'$ which characterizes the vanilla QUBO model (\ref{equation:vanilla_qubo_model}). It is an upper triangular matrix, all of whose entries are non-zero. The number of non-zero entries of $\hat{Q}'$ is at most 
\begin{equation}
    \# \text{non-zero entries of } \hat{Q}' \leq \frac{12 \times (12+1)}{2} = 78.
\end{equation}
The matrix congruence relation reduces $\hat{Q}'$ to a block diagonal matrix $\hat{Q}$ from (\ref{equation:Qhat_y}), comprised of $4$ upper triangular block matrices of dimension $3 \times 3$. The number of non-zero entries of the matrix $\hat{Q}$ characterizing the new QUBO model, as in (\ref{equation:new_qubo_model}), is at most
\begin{equation}
    \# \text{non-zero entries of } \hat{Q} \leq 4 \times \frac{3 \times (3+1)}{2} = 24.
\end{equation}
Indeed, $\hat{Q}$, as a block diagonal matrix comprised of $4$ upper triangular matrices of size $3 \times 3$, has all but one non-zero entries. The matrix congruence relation cuts down the number of non-zero entries of the characterizing matrix of the QUBO model by more than a factor of $\frac{1}{3}$:
\begin{equation}
    \frac{\# \text{non-zero entries of } \hat{Q}}{\# \text{non-zero entries of } \hat{Q}'} \leq \frac{24}{78} < \frac{1}{3}.
\end{equation}

The results thus verify that exploiting congruence relation between symmetric and diagonal matrices substantially simplifies and enhances the QUBO model for solving systems of linear equations. As for determining the matrices $D$ and $R$ from (\ref{equation:sylvester_final}), there is room for improvement on solving systems of linear equations cost-efficiently by utilizing advantages quantum computing methods possess that classical algorithms do not. For example, some classical implementations such as QR or SVD decomposition are not effectively parallelizable or computationally expensive. Meanwhile, distinctive features of quantum computing methods, such as computational bases \cite{MLZ20} or synthesis of quantum circuits \cite{BBVA20}, can contribute to computationally cost-efficient QR decomposition algorithms. We shall thus expect to achieve polynomial speedup in solving systems of linear equations by combining quantum annealing approaches.

\section*{Declaration of Interests}
The authors have no competing interests which may have influenced the work shown in this manuscript.

\section*{Acknowledgements}
This work was partially supported by National Institute for Mathematical Sciences (NIMS) grant funded by the Korean Government (MSIT) B21810000.

\end{document}